\documentclass[12pt]{iopart}

\usepackage{graphicx}
\usepackage{setstack}
\usepackage{iopams}
\usepackage{color}

\begin{document}

\title[First passage times for two-channel diffusion]{First passage time statistics for two-channel diffusion}

\author{Alja\v{z} Godec$^{1,2}$\footnote{agodec@uni-potsdam.de} and Ralf Metzler$^{1}$}
\address{$1$ Institute of Physics \& Astronomy, University of Potsdam, 14476
Potsdam-Golm, Germany\\
$2$ Department of Molecular Modeling, National Institute of Chemistry, 1000 Ljubljana, Slovenia}

\begin{abstract}
We present rigorous results for the mean first passage time and first passage time statistics for two-channel Markov additive diffusion in a 3-dimensional spherical
domain. Inspired by biophysical examples we assume that the particle can only recognise the
target in one of the modes, which is shown to effect a non-trivial first passage 
behaviour. We also address the scenario of intermittent
immobilisation. In both cases we prove that despite the perfectly non-recurrent motion
of two-channel Markov additive diffusion in  3 dimensions the first passage
statistics at long times do not display Poisson-like behaviour if none
of the phases has a vanishing diffusion coefficient. This stands in
stark contrast to the standard (one-channel)
Markov diffusion
counterpart. We also discuss the relevance of our results in the
context of cellular signalling.
\end{abstract}
 
\section{Introduction}

When does a stochastic variable reach a preset threshold (e.g., a
physical target or a given asset value) for
the first time? This generic first passage time (FPT)
problem \cite{Sid,Ralf} is
central to the kinetics across many disciplines, such as diffusion controlled
chemical reactions \cite{Smol}, signalling cascades in biological cells
\cite{Alberts,Holcman1,Holcman2,Holcman,otto,GodSciRep,GodPRE}, transport in disordered media \cite{bAv} including the
breakthrough dynamics in hydrological aquifers \cite{brian}, the location of
food by foraging bacteria and animals \cite{foraging}, up to the global spreading
of diseases \cite{Lloyd,dirkle} or stock market dynamics \cite{mantegna}.

Despite their diverse phenomenology and owing to the central limit theorem, the kinetics in stochastic systems such as
the above can often
be mapped onto a standard Markovian random walk. Here we will discuss
the FPT behaviour in the context of a particle diffusing in space. In open domains the FPT statistics of the random walk---or its diffusion
limit---decay as a power law\footnote{with a logarithmic correction in
dimension 2}, giving rise to a diverging mean FPT (MFPT)
\cite{Sid}. Heavy tails are common when it comes to persistence properties of infinite
systems \cite{Bray}. A finite domain size, however, introduces an exponential long time decay and thus a finite MFPT, which becomes
a function of the system size and dimensionality
\cite{Sid,GodPRE,GodSciRep,Carlos,Olivier}. 
In unbounded domains all first passage trajectories are
nominally \emph{direct} \cite{GodSciRep}, whereas in confinement a
particle can arrive at the target site also via reflection with the
confining boundary, i.e., via an \emph{indirect} trajectory \cite{GodSciRep}.
Moreover, the MFPT for
non-recurrent and translation invariant Markov dynamics is often strongly dominated
by the long time behaviour--by \emph{indirect} trajectories
\cite{GodSciRep,Olivier}. This is the case when the volume of the
domain tends to be very large and/or the target size tends to be very small \cite{Olivier}.
In such non-recurrent scenarios knowing the MFPT fully--yet non-trivially--
characterises the long time asymptotic of the FPT statistics
\cite{GodSciRep,Olivier}. In non-recurrent systems with strongly broken
translation invariance an additional time scale emerges, mirroring brief
excursions away from the target \cite{GodSciRep}. This intermediate
time scale in turn significantly contributes to the MFPT
\cite{GodSciRep}. Conversely, for recurrent motion
the rate of the long time exponential decay is strongly affected by both, direct
and indirect trajectories \cite{Olivier}. 

Often the dynamics additionally depend on some internal state, such as for example in the
so-called 'intermittent search model', where the particle
randomly switches between passive diffusion and active ballistic
motion in a Poissonian \cite{Gleb,OlivierAct,GodecActive,GodecJPA} or
L\'evy \cite{Koren} fashion, or equivalently in the 'facilitated diffusion model of gene regulation'
\cite{gene}, where the particle switches between 3-dimensional and
1-dimensional diffusion, with an additional dynamical component due to
conformational dynamics of DNA \cite{Wuite}, which in the annealed
limit gives rise to L\'evy flights \cite{Lomholt}. A similar case is
the transitioning between search and recognition modes in the
1-dimensional search of transcription factors along DNA
\cite{SlutskyBauer}.

A similar random transitioning occurs in
stochastically gated chemical reactions \cite{Holcman,Greact} and
stochastically gated narrow escape \cite{Holcman2,Holcman}. The
Markovian switching between the internal states introduces a much richer
phenomenology and can lead to qualitative changes in the FPT
statistics, such as in the case of the random search for a stochastically gated target
\cite{Greact}. Conversely, by combining recurrent and non-recurrent
motion phases and thereby suppressing oversampling on large spatial
scales and improving the hitting on small length-scales, one can
improve stochastic search processes in the sense of minimising the
MFPT to reach the target \cite{Gleb,OlivierAct,gene}. 

From a
mathematical point of view all these compound processes are called \emph{Markov
  additive} (MA) \cite{MarkovAdd}. MA processes are a class of Markov
processes, whose state space $G=\Omega\times F$ is at least 2-dimensional and can be
split into $\Omega$, a Markovian component and an additive component $F$, which is translation invariant \cite{MarkovAdd}.  
Formally, some features of the FPT properties of MA
processes with a general state space have already been addressed in
the mathematical literature using algebraic methods (see, 
e.g., \cite{MAddFPT}). Yet, explicit results on the FPT statistics for MA
processes are sparse. Moreover, the
interplay between (non)recurrent motion and Markovian switching between
internal states and its comparison to standard Markov diffusion
processes remains elusive.  

Here we present rigorous results for the MFPT and
FPT statistics for two-channel Markov additive diffusion\footnote{Note that the term 'double
  diffusion' also appears in the literature \cite{Hughes}.} in a 3-dimensional spherical
domain with the additive component $F$ being Markovian. We consider a gated particle, that is, the particle can only recognise the
target in one of the modes, which is shown to lead to non-trivial FPT
behaviour. We also address the FPT problem of transitioning to an
immobile phase. In
particular, we prove that despite the perfectly non-recurrent motion
of two-channel Markov additive diffusion in 3 dimensions,
the MFPT does not fully specify the asymptotic exponential decay of
the FPT statistics as soon as none of the phases is static (i.e., has
a vanishing diffusion coefficient), in contrast to the standard Markovian
counterpart.  
  
The paper is organised as follows. First we set up the model of
two-channel MA as a mixed boundary value problem for two coupled
forward Fokker-Planck equations. Next, we summarise our main
results on the MFPT and FPT statistics and discuss the implications of our results in a biophysical context. 
In the following sections we
present detailed calculations, proofs and additional
technicalities. As these contain essential mathematical approaches we
include here all crucial steps of the derivation. Finally, we give a concluding perspective and discuss
possible extensions of the work.

\section{Markov additive two-channel diffusion}

\begin{figure}
\begin{center}
\includegraphics[width=7.cm]{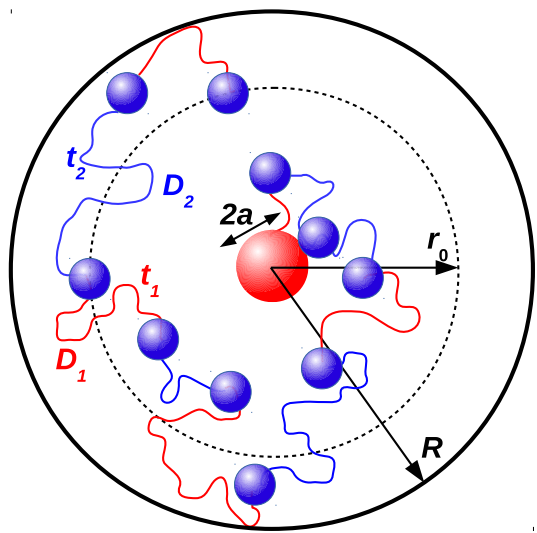}
\caption{Schematic of the model system: A particle performs 3$d$
  Brownian motion in a spherical domain with a reflecting confining
  boundary at $R$ and randomly switches between diffusion coefficients
$D_1$ and $D_2$. The duration of the respective phases is
  exponentially distributed with rates $k_{1,2}$. The particle only
  recognises the target (red sphere in the centre) in phase $1$ -- the
  recognition mode, whereas it experiences the target as a reflecting
sphere in phase $2$ -- the
  non-recognition mode.}
\label{schm}
\end{center}
\end{figure}

We consider a 3-dimensional
spherical domain of size $R$ with a perfectly absorbing target with
radius $a$ at the centre (see Fig.~\ref{schm}). The particle's
diffusion coefficient $D_k$, i.e., the internal variable, randomly switches between states $k=1$ and $k=2$ in
a Markovian fashion with rates $k_1$ and $k_2$, respectively. In other words,  the
duration of the respective phases is exponentially distributed with
mean $\langle \tau_1\rangle = k_1^{-1}$ and $\langle
\tau_2\rangle=k_2^{-1}$. 
At any instance, the particle's dynamics on
infinitesimal time-scales $\Delta t$ can be discretized as (see e.g. \cite{Holcman,Schuss}) 
\begin{equation}
\label{discrete}
\mathbf{x}_i(t+\Delta t)=\left\{
\begin{array}{rl}
\mathbf{x}_j(t) & \mathrm{w.p.}\quad  k_i\Delta t,\\
\mathbf{x}_i(t) + \sqrt{2D_i\Delta t}\boldsymbol{\xi}(t) & \mathrm{w.p.}\quad 1-k_i\Delta t.
\end{array} \right. 
\end{equation}
where $\mathrm{w.p.}$ denotes 'with probability' and with $i\ne j=1,2$
and $\xi_k$ being the component of a zero mean Gaussian white noise with $\langle
\xi_k(t)\xi_l(t')\rangle=\delta(t-t')\delta_{kl}$\footnote{Here
  $\delta(x)$ and $\delta_{ik}$ denote the Dirac and Kroenecker delta
  functions, respectively.}. We introduce the
propagator $p(\mathbf{x},t;i|\mathbf{y},t';j)$ -- the transition
probability density for the particle to be at $\mathbf{x}$ at time $t$ in internal
state $i$ given that it was previously at $\mathbf{y}$ at time $t'$ in
the internal
state $j$. To first order in $\tau$ during any infinitesimal interval $\tau=t-t'$ the
propagator can be split into two steps, (i) switching from $j$ to the internal
state $i$ without diffusion, and (ii) diffusion in this new state without
switching:
\begin{eqnarray}
\label{generator}
p(\mathbf{x},t;i|\mathbf{y},t';j)=\left[k_j\tau(1-\delta_{ij})+\delta_{ij}(1-k_j\tau)(1+D_j\nabla^2)\right]\delta(\mathbf{x}-\mathbf{y})\nonumber\\
=\left[(1-k_j\tau)\delta_{ij}+(k_j(1-\delta_{ij})+\delta_{ij}D_j\nabla^2)\tau\right]\delta(\mathbf{x}-\mathbf{y})+\mathcal{O}(\tau^2), 
\end{eqnarray}      
where $\nabla^2$ is taken with respect to $\mathbf{x}$.
Using Eq.~(\ref{generator}) as well as the Chapman-Kolmogorov
equation \cite{Holcman,Schuss}
\begin{equation}
\label{ChKol}
p(\mathbf{x},t+\tau;i|\mathbf{y},t';j)=\int_{\Omega}d\mathbf{z}\sum_{k=1}^2p(\mathbf{x},t+\tau;i|\mathbf{z},t;k)p(\mathbf{z},t;k|\mathbf{y},t';j),
\end{equation}
we obtain, upon taking the limit $\tau\to 0$ the
forward Fokker-Planck equation (FPE), which for convenience we write in a
vector form as
\begin{equation}
\label{FPE}
\partial_t \mathbf{p}^{\mathrm{T}}(\mathbf{x},t|\mathbf{y},t')=\left(
\begin{array}{cc}
D_1\nabla^2 - k_1 & k_2 \\
k1 & D_2\nabla^2 - k_2 
\end{array} \right)\mathbf{p}^\mathrm{T}(\mathbf{x},t|\mathbf{y},t'),
\end{equation}
where $\mathbf{p}=(p_1,p_2)$ is the transition
probability density vector with the general initial condition
$\mathbf{p}(\mathbf{x},0)=(w\delta(\mathbf{x}-\mathbf{x}_{0,1}),[1-w]\delta(\mathbf{x}-\mathbf{x}_{0,2}))$
with arbitrary real $w\in[0,1]$. As the system is linear the solution
to this general initial condition can be reconstructed from the
solutions for $w=1,0$ and
$\mathbf{x}_{0,1}=\mathbf{x}_{0,2}=\mathbf{x}_{0}$. The FPE
(\ref{FPE}) is complemented by inhomogeneous boundary conditions at
the surface of the target and confining boundary, $\partial\Omega_a$ and $\partial\Omega_R$, respectively: 
\begin{eqnarray}
\label{BC}
\left.p_1(\mathbf{x},t)=\nabla p_2(\mathbf{x},t)\cdot\hat{\mathbf{n}}_a\right|_{\mathbf{x}=\partial\Omega_a}=0,\nonumber\\ 
\left.\nabla p_1(\mathbf{x},t)\cdot\hat{\mathbf{n}}_R=\nabla p_2(\mathbf{x},t)\cdot\hat{\mathbf{n}}_R\right|_{\mathbf{x}=\partial\Omega_R}=0,
\end{eqnarray} 
where $\hat{\mathbf{n}}_a$ and $\hat{\mathbf{n}}_R$ denote the
respective surface normals.  
The FPT probability density is obtained from the flux into the
absorbing target from the recognition phase $1$
\begin{equation}
\label{flux}
\left.\wp(t)=4\pi a^2D_1\nabla p_1(\mathbf{x},t)\cdot\hat{\mathbf{n}}_a\right|_{\mathbf{x}=\partial\Omega_a}
\end{equation}  
and the MFPT corresponds to the first moment of $\wp(t)$, namely
$\langle t\rangle=\int_0^{\infty}t\wp(t)dt$. All quantities are made
dimensionless by expressing time in units of $\tau_0=R^2/D_1$, length,
or in fact radii, in units of the domain radius $r_i\to x_i\equiv r_i/R$ and by
introducing the dimensionless ratios $z=k_2/k_1$ and
$\varphi=D_2/D_1$. Note that the time unit $\tau_0$ is 'natural' as it holds
trivially for any normal diffusion process with a hyperspherical
symmetry that the mean first passage
time scales as $\langle t\rangle\propto R^2$ with the confining
hypersphere radius $R$, irrespective of the dimension \cite{Hughes}.

In the Brownian Dynamics
simulations reported herein the dynamics are implemented by first drawing a sojourn
time $\tau_s$
from the respective exponential density with mean $k_{1,2}^{-1}$ and
then propagating the particle's position within the interval $\tau_s$
according to the overdamped
Langevin equation with the respective diffusion coefficient
$D_{1,2}$. The initial condition is sampled uniformly over the surface
of a sphere with radius $r_0$. Reflecting boundary conditions are implemented by
neglecting any move that would take the particle into the reflecting
boundary (while still updating the time). The particle is
propagated until it reaches the target while being in the recognition
mode $1$.  

\section{Summary and discussion of the main results}

\subsection{Mean first passage times}
We first focus on the MFPT. The proofs of the equations presented in
this sections will be described in later sections. The MFPT to arrive
at $x_a$ if
starting from $x_0$ in the recognition mode 1, $\langle t_{x_a}(x_0)\rangle_1$, is given exactly as
\begin{equation}
\label{MFPT1}
\langle t_{x_a}(x_0)\rangle_1=\frac{1+z}{z+\varphi}\left[\langle
t_{x_a}(x_0)\rangle_0+\frac{\varphi(1-x_a^3)}{3z\overline{\mathcal{D}}_1(x_a)}\left(\frac{\overline{\Delta}_1(x_a)}{x_a}-\frac{\overline{\Delta}_1(x_0)}{x_0}\right)\right],
\end{equation} 
where we introduced the mean first passage time of standard 3-dimensional Brownian
motion 
\begin{equation}
\label{MFPT0}
\langle
t_{x_a}(x_0)\rangle_0=\frac{1}{3}\left(x_a^{-1}-x_0^{-1}-\frac{x_0^2-x_a^2}{2}\right)
\end{equation} 
 as well
as the auxiliary functions 
\begin{eqnarray}
\label{aux1}
\overline{\mathcal{D}}_1(y)&=&(1-y)\cosh[\sqrt{1+z/\varphi}(1-y)]\nonumber\\
&&+\frac{[y(1+z/\varphi)-1]\sinh[\sqrt{1+z/\varphi}(1-y)]}{\sqrt{1+z/\varphi}}\\
\label{aux2}
\overline{\Delta}_1(y)&=&\cosh[\sqrt{1+z/\varphi}(1-y)]-\frac{\sinh[\sqrt{1+z/\varphi}(1-y)]}{\sqrt{1+z/\varphi}}.
\end{eqnarray}
Note that the prefactor in Eq.~(\ref{MFPT1}) is just the
inverse of the effective diffusion coefficient
$D_{\mathrm{eff}}=(D_1/k_1+D_2/k_2)/(k_1^{-1}+k_2^{-1})$ expressed in
units of $D_1$. Note that if the switching between the internal states
is fast compared to the
time needed to arrive to the vicinity of the target, then trajectories
essentially behave as 3-dimensional Brownian motion with an effective diffusion coefficient
$D_{\mathrm{eff}}$. Thus $\langle t_{x_a}(x_0)\rangle_1$ has the form of
the MFPT of standard 3-dimensional Brownian motion with  $D_{\mathrm{eff}}$ plus a term compensating for
the contribution of trajectories where the particle does not switch
between modes sufficiently many times.   

\begin{figure}
\begin{center}
\includegraphics[width=14.cm]{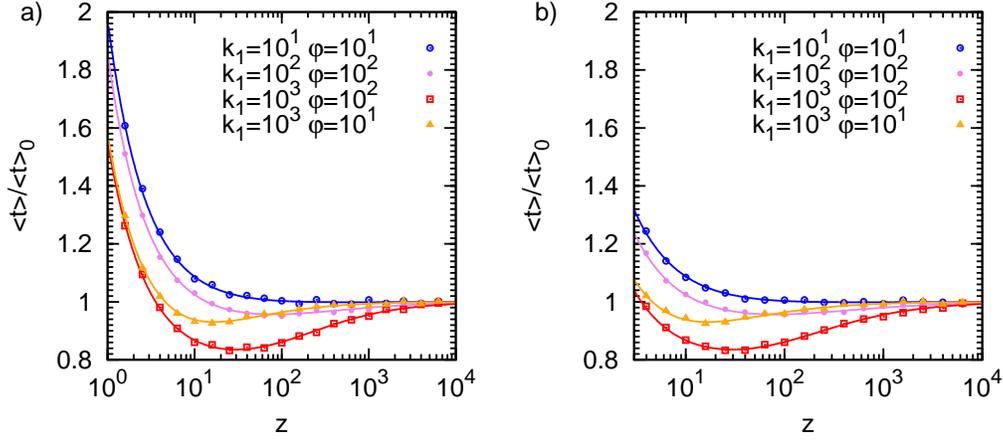}
\caption{Normalised MFPT as a function of $z=k_2/k_1$ for various values of $k_1$
  and $\varphi$. The results correspond to $x_0=0.5$ and
  $x_a=0.01$. Panel a) corresponds to a particle starting in the
  recognition mode $1$, whereas b) depicts the results for starting
  in the non-recognition mode $2$.}
\label{MFPTs}
\end{center}
\end{figure}

The result in Eq.~(\ref{MFPT1}) as a function of $z$ for various
values of $k_1$ and $\varphi$, divided by $\langle t_{x_a}(x_0)\rangle_0$, is depicted in Fig.~\ref{MFPTs}a) (full lines) and shows
excellent agreement with Brownian Dynamics simulations (symbols). Note
that intuitively for sufficiently large $\varphi=D_2/D_1$, the MFPT $\langle
t_{x_a}(x_0)\rangle_1$ can be significantly shorter than $\langle
t_{x_a}(x_0)\rangle_0$. In addition, for sufficiently large $k_1$ there
exists an optimal value of $z$ where $\langle
t_{x_a}(x_0)\rangle_1$ has a minimum. The optimisation of $\langle
t_{x_a}(x_0)\rangle_1$ , which essentially corresponds to solving a
non-linear algebraic equation for $z$, is beyond the scope of the present work.    

\begin{figure}
\begin{center}
\includegraphics[width=16.cm]{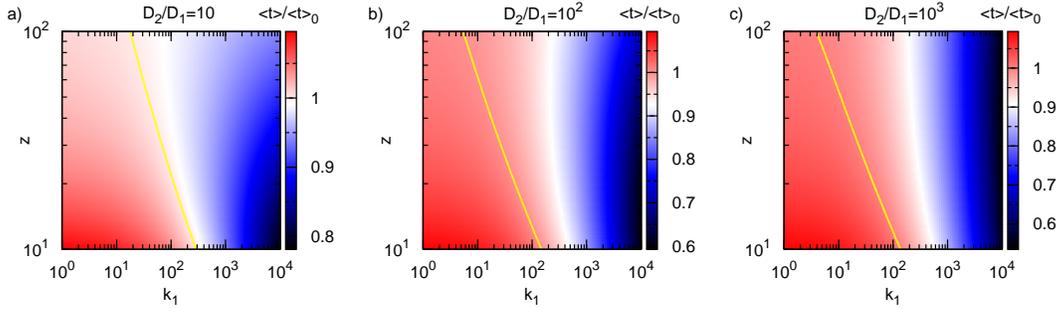}
\caption{Normalised MFPT $\langle
t_{x_a}(x_0)\rangle_1/\langle
t_{x_a}(x_0)\rangle_0$ as a function of $k_1$ and $z=k_2/k_1$ for
various $\varphi$ and $x_0=0.5$ and
  $x_a=0.01$. The yellow contour corresponds to $\langle
t_{x_a}(x_0)\rangle_1/\langle
t_{x_a}(x_0)\rangle_0=1$.}
\label{MFPTr}
\end{center}
\end{figure}

Additional insight is obtained from the joint dependence of $\langle
t_{x_a}(x_0)\rangle_1$ on $z$ and $k_1$. The results for three
different values of $\varphi$ are shown in Fig.~\ref{MFPTr}.
As already mentioned, for sufficiently large $\varphi$ we find that $\langle
t_{x_a}(x_0)\rangle_1$ intuitively decays with increasing $k_1$, as it
is beneficial if the particle spends more time in the faster diffusing
phase. The dependence on $z$ is however, non-monotonic due to the simple
fact that there is always a trade-off between reaching the target in
the (faster) non-recognition mode and hitting the target from close
distance through the recognition mode. For example, for
large $\varphi$, the effective $D_{\mathrm{eff}}$ can become very large and
consequently the MFPT can decrease substantially as long as
$z$ is not too close to $\varphi$, i.e. the particle spends enough time in the non-recognition
mode. Intuitively, for $z\to\infty$ and $\varphi$ finite
$D_{\mathrm{eff}}$ converges to 1. However, if $z\to 0$, i.e., very long
residence time in the non-recognition mode, the second term of
Eq.~(\ref{MFPT1}) diverges as $1/z$ because even if the motion in mode
2 is fast enough to essentially reach a local steady state, the rate
to switch back to the recognition mode becomes rate limiting.    

Conversely, $\langle t_{x_a}(x_0)\rangle_2$ , the MFPT to $x_a$
starting from $x_0$ in the non-recognition mode $2$ is given exactly as
\begin{equation}
\label{MFPT2}
\langle
t_{x_a}(x_0)\rangle_2=t^h_{\mathrm{eff}}+\frac{1+z}{z+\varphi}\left[\langle
t_{x_a}(x_0)\rangle_0+\frac{(1-x_a^3)}{3\overline{\mathcal{D}}_1(x_a)}\left(\frac{\varphi}{z}\frac{\overline{\Delta}_1(x_a)}{x_a}+\frac{\overline{\Delta}_1(x_0)}{x_0}\right)\right],
\end{equation}
where we introduced the effective time to hit the target from the
non-recognition mode $2$ once arriving within a distance to the
target, which corresponds to the typical distance
moved in a switching cycle $k_1^{-1}+k_2^{-1}$
\begin{equation}
\label{effrate}
t^h_{\mathrm{eff}}=k_1^{-1}\frac{(D_1-D_2)/k_2}{D_1/k_1+D_2/k_2}\equiv
k_1^{-1}\frac{1-\varphi}{z+\varphi}.
\end{equation}
Note that the effective hitting-time correction $t^h_{\mathrm{eff}}$ can be positive or negative depending on $\varphi$ .
The result in Eq.~(\ref{MFPT2}) as a function of $z$ for various
values of $k_1$ and $\varphi$, expressed relative to $\langle
t_{x_a}(x_0)\rangle_0$, is depicted in Fig.~\ref{MFPTs}b) (full lines)
and as before shows
excellent agreement with Brownian Dynamics simulations (symbols). 
\begin{figure}
\begin{center}
\includegraphics[width=16.cm]{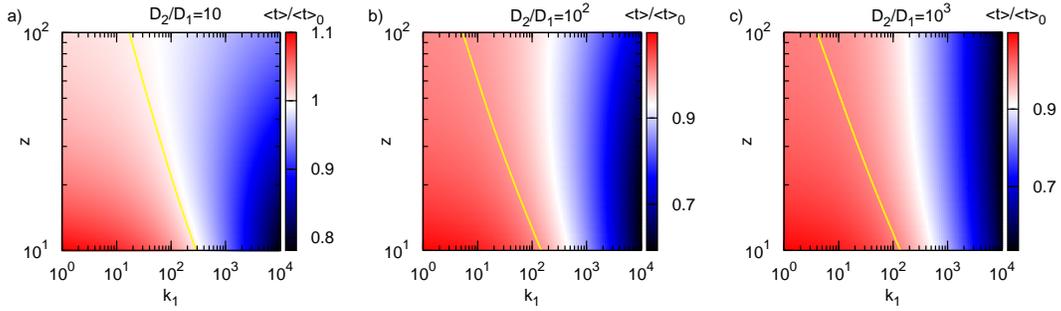}
\caption{Normalised MFPT $\langle
t_{x_a}(x_0)\rangle_2/\langle
t_{x_a}(x_0)\rangle_0$ as a function of $k_1$ and $z=k_2/k_1$ for
various $\varphi$ and $x_0=0.5$ and
  $x_a=0.01$. The yellow contour corresponds to $\langle
t_{x_a}(x_0)\rangle_0/\langle
t_{x_a}(x_0)\rangle_2=1$.}
\label{MFPTnr}
\end{center}
\end{figure}
Qualitatively, the scenario of starting in the non-recognition mode is
very similar to the previous one. 

To understand the subtle difference between the two initial conditions more
deeply we inspect the meaning of the effective hitting-time correction
$t^h_{\mathrm{eff}}$ in Eq.~(\ref{effrate}) in more
detail. If $\varphi\ll 1$ then $t^h_{\mathrm{eff}}\sim 1/k_2$,
i.e., the correction time is equal to the mean time spent in the non-recognition mode. Conversely, $t^h_{\mathrm{eff}}$ gives a large negative contribution to $\langle
t_{x_a}(x_0)\rangle_2$ when $k_1$ is small and $\varphi\gg1$ and
$\varphi\gg z$, that is, the particle resides over long periods in the recognition
mode while simultaneously the typical distance moved in mode 2 is much
larger than the one moved in mode 1, $D_2/k_2\gg D_1/k_1$. The
particle therefore has no difficulty in hitting the target from mode 2
as soon as it arrives to within a typical distance to it. However, as
this also implies a small $D_1$ the natural time unit $\tau_0=R^2/D_1$
explodes and the search time increases. 

Finally, we inspect the scenario of intermittent immobilisation such
as occurring in chromatography, i.e., $\varphi=0$. We find from
Eqs.~(\ref{MFPT1}) and (\ref{MFPT2}) that
\begin{eqnarray}
\label{MFPTstat1}
\langle t_{x_a}(x_0)\rangle_{1,s}&=&(1+z^{-1})\langle
t_{x_a}(x_0)\rangle_0,\nonumber\\
\langle t_{x_a}(x_0)\rangle_{2,s}&=&(1+z^{-1})\langle
t_{x_a}(x_0)\rangle_0+k_2^{-1}.
\label{MFPTstat2}
\end{eqnarray}
Note that the prefactor $1+z^{-1}$ is the inverse of the steady-state
probability to be in the recognition mode 1. The results in
Eq.~(\ref{MFPTstat2}) are intuitive as the
diffusion coefficient becomes trivially reduced by the fraction of
time spent in mode 1 (since mode 2 is static). Moreover, $\langle
t_{x_a}(x_0)\rangle_{2,s}$ contains the additional term accounting for
the fact that the particle needs to switch to mode 1 exactly once more
since it starts from mode 2.   

\subsection{Probability density of first passage times}

Due to the complexity of the problem it is not possible to obtain a general exact closed-form expression
for $\wp(t)$ valid on all time scales. Therefore we here limit the
discussion to the exact long time asymptotic of $\wp(t)$. In this
section we simply state the results, whereas the proofs are presented
in the next section. 
As
intuitively expected (and proven in section 4.3) all moments of
$\wp(t)$ are finite as longs as $x_a>0$ and $R<\infty$. Moreover, as
$\wp(t)$ is smooth, this implies that it decays exponentially for long
times, $\wp(t)\sim\mathcal{C}(x_0,x_a)\mathrm{e}^{-\lambda_0t}$, where $\sim$ stands for asymptotic equality. Exact
expressions for $\lambda_0$ and $\mathcal{C}(x_0,x_a)$ can be obtained
from $\tilde{\wp}(s)$, the Laplace transform of the FPT density
$\wp(t)=\hat{\mathcal{L}}^{-1}\{\tilde{\wp}(s)\}$. The results
read  
\begin{eqnarray}
\label{full}
\lambda_0(x_a)&=&\sum_{k=1}^{\infty}\frac{v^{(0)}(x_a)^k}{v^{(1)}(x_a)^{2k-1}}\frac{\mathrm{det}\mathcal{M}_k}{(k-1)!}\\
\mathcal{C}(x_0,x_a)&=&\lim_{k\to\infty}\frac{\sum_{l=0}^{k-1}\left[u^{(l)}(x_0)-v^{(l)}(x_a)u^{(k)}(x_0)/v^{(k)}(x_a)\right](-\lambda_0)^l/l!}{\sum_{l=0}^{k-1}\sum_{m=1}^{k-l}v^{(l+m)}(x_a)(-\lambda_0)^{l+m-1}/(l+m)!},
\label{pref}
\end{eqnarray}
where $u^{(k)}(x_0)$ and $v^{(k)}(x_a)$ denote the $k$th order
derivative of the numerator and denominator of $\tilde{\wp}(s)$ with
respect to $s$, respectively, evaluated at $s=0$ (defined in section 4.3) and $\mathcal{M}_k$
stands for the 'almost' triangular matrix with elements
\begin{eqnarray}
\label{matrix}
\mathcal{M}_k(i,j)=&&\frac{v^{(i-j+2)}\Theta(i-j+1)}{(i-j+2)!}\nonumber\\
&&\times\left[k(i-j+1)\Theta(j-2)+i\Theta(1-j)+j-1\right],
\end{eqnarray}
where $\Theta(n)$ denotes the discrete Heaviside step function and
with the symbolic convention $\mathrm{det}\mathcal{M}_1\equiv 1$. Note
that Eqs.~(\ref{full}) and (\ref{matrix}) are fully general and are derived under very mild
assumptions, which are warranted by the physics of the problem. More
precisely, one has to assume
(i) that all moments of $\wp(t)$ exist, (ii) that
$\tilde{\wp}(s)$ has no branch points on the negative real axis, and
(iii) that $\lim_{k\to\infty}u^{(k)}/v^{(k)}<\infty$. While (i) is satisfied
trivially, (ii)\footnote{One can show for most Markov processes,
  incl. Brownian motion (BM) in dimensions 1, 2, and 3, diffusion on fractals, uniformly biased 1-dimensional
  BM, radially biased 2-dimensional BM and the
  Ornstein-Uhlenbeck process, that $\tilde{\wp}(s)$ has only simple poles and
  removable singularities on the negative real axis \cite{Godec_PRX}.} and (iii) are borne out in practice (see section 4.3).   
 
We are particularly interested in the physically relevant scenario of
a small target size. In the present case Eq.~(\ref{full}) actually defines a power series in $x_a$
and we find in the limit $x_a\ll 1$ (note that for convenience we here
present the inverse of $\lambda_0$) 
\begin{equation}
\label{rateR}
\lambda_0^{-1}(x_a)= \langle
t_{x_a}(1)\rangle_2-\left[\frac{1+z}{z+\varphi}\right]\frac{1}{3\overline{\mathcal{D}}_1(x_a)}+\left[\frac{z/\varphi+\varphi}{z+\varphi}\right]\frac{\overline{\mathcal{D}'}_1(x_a)}{\overline{\mathcal{D}}_1(x_a)}+\mathcal{O}(x_a),
\end{equation}
where we introduced the auxiliary function 
\begin{eqnarray}
\label{AuxD}
\overline{\mathcal{D}'}_1(y)&=&\frac{(1-y)[(1+z/\varphi)y-1]\cosh[\sqrt{1+z/\varphi}(1-y)]}{(1+z/\varphi)}\nonumber\\
&&+\frac{[(1+z/\varphi)(1-y+y^2)+1]\sinh[\sqrt{1+z/\varphi}(1-y)]}{(1+z/\varphi)^{3/2}}.
\end{eqnarray}
Analogously, the series in (\ref{pref}) converges with the first term
for $x_a\to 0$ and we obtain the exact asymptotic result
\begin{equation}
\label{FPTas}
\wp_{1,2}(t)\sim\langle t_{x_a}(x_0)\rangle_{1,2} \lambda^2_0(x_a)\mathrm{e}^{-\lambda_0(x_a)t}. 
\end{equation}
Eq.~(\ref{FPTas}) is the central result of this
paper. It reveals that the exponential decay rate is independent of
the initial condition (i.e. the position as well as the internal
state). This regime describes indirect trajectories, which interact
with the confining boundary before heading towards the target \cite{GodSciRep}. The
fact that the decay rate of $\wp_{1,2}(t)$  is independent of the
initial condition suggests that reaching the external boundary from
the initial location is much faster
than reaching the target from the external boundary. Moreover, only
the prefactor depends on the initial condition--the position as well as internal
state, which suggests the statistics of direct trajectories, i.e., those that reach
the target without ever interacting with the boundary, controls the
statistical weight of the exponential asymptotic, which is equivalent
to the simpler Markovian
counterparts \cite{GodSciRep}. To see this we can rewrite Eq.~(\ref{FPTas})
as a product of the 'weight' factor $\langle t_{x_a}(x_0)\rangle_{1,2}\lambda_0(x_a)$ and a normalised exponential
$\mathrm{e}^{-\lambda_0(x_a)t}\lambda_0(x_a)$. Therefore, the
contribution of the long-time regime to 
expectations taken over $\wp_{1,2}(t)$ will depend only on the
'weight' factor and the smallest time where  Eq.~(\ref{FPTas}) becomes
valid.  

Moreover,  Eq.~(\ref{FPTas}) highlights that the asymptotic of
$\wp(t)$ \emph{cannot} be fully reconstructed by knowing the MFPT, as both
the prefactor and the exponent contain a non-trivial correction term
in $\lambda_0$. This observation is in stark contrast to the simpler Markovian
counterpart, where  the $\wp(t)$
asymptotic can indeed be reconstructed once the MFPT is known (see
\cite{GodSciRep,Olivier}) as long as the dynamics is
non-recurrent, highlighting the non-trivial first passage character
of Markov additive processes. 

\begin{figure}
\begin{center}
\includegraphics[width=\textwidth]{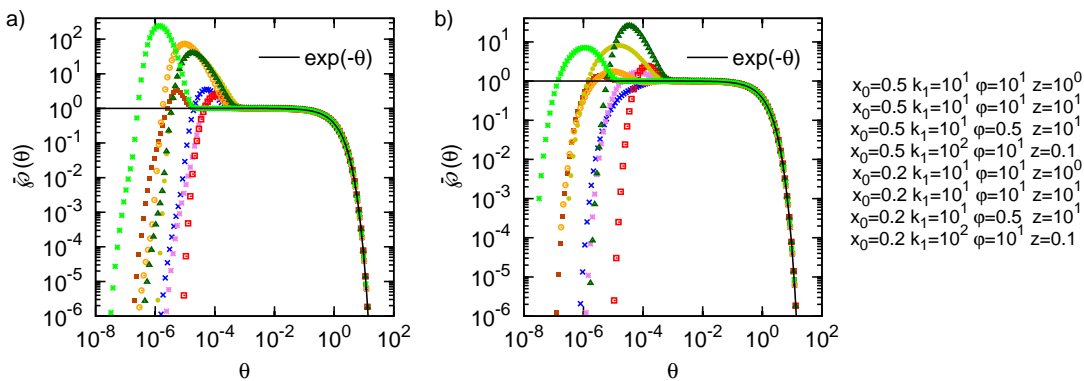}
\caption{Rescaled FPT probability
  density $\overline{\wp}_{1,2}(\theta)\equiv \wp_{1,2}(\theta)/(\lambda^2_0(x_a)\langle
t_{x_a}(x_0)\rangle_{1,2})$ obtained from Brownian dynamics simulations as a function of
  the rescaled time $\theta=t/\lambda_0$ for $x_a=0.01$ and
  various $k_1,z$ and $\varphi$ and two different initial conditions
  $x_0$ for the scenario of a) starting in the recognition mode and b)
  starting in the non-recognition mode. The full black
line corresponds to the unit exponential master scaling in Eq.~(\ref{masterC}). The simulation results perfectly collapse on the master curve.}
\label{FPTden}
\end{center}
\end{figure}

Furthermore, if we rescale time according to $\theta\equiv
t/\lambda_0(x_a)$, then all FPT densities must collapse onto the
master curve 
\begin{equation}
\label{masterC}
\overline{\wp}_{1,2}(\theta)\equiv \wp_{1,2}(\theta)/[\lambda_0^2(x_a)\langle
t_{x_a}(x_0)\rangle_{1,2}]=\mathrm{e}^{-\theta}.
\end{equation} 
Indeed, this collapse
is shown in Fig.~\ref{FPTden} for a variety of parameters and initial
conditions. 

In the case of a static non-recognition mode we again find for
$x_a\ll 1$ an
intuitive renormalisation of the diffusion coefficient and we can
identify the
universal form for non-recurrent single-channel Markov dynamics \cite{GodSciRep}
\begin{equation}
\label{FPTasS}
\wp_{1,2}^s(t)\sim\langle
t_{x_a}(x_0)\rangle_{1,2;s} \langle t_{x_a}(1)\rangle_{2,s}^{-2}\mathrm{e}^{-t/\langle t_{x_a}(1)\rangle_{2,s}}. 
\end{equation}
Note that in contrast to $\varphi\ne 0$ (see Eq.~(\ref{FPTas})), the
transient immobilisation case leads to a Poisson-like asymptotic (\ref{FPTasS}) \cite{GodSciRep,Olivier}.

\subsection{Biophysical implications of the results}

The stochastic switching between different internal states is
relevant in various biophysical problems, in particular in cellular
signalling pathways. Namely, proteins can
switch between different conformations with different diffusivities, either spontaneously or upon
interaction with other signalling molecules
\cite{Alberts,Holcman,Greact}. Similarly, in the regulation of gene transcription regulatory
proteins can change the affinity of TF for the promoter
site \cite{Berg2,Maj}. Most proteins transiently bind non-specifically to other
proteins and other cytoplasmic constituents, incl. immobilised
structures \cite{Verkman,CAM}.
Furthermore, some signalling molecules such as
calmodulin (a cellular calcium sensor) are intrinsically 'sticky' and
bind to various cytoplasmic constituents when biochemically stimulated
(in the case of calmodulin by calcium \cite{CAM}), and as a result
display a smaller diffusion coefficient in the activated mode. In these cases only
the active form typically binds to its target and triggers a
biological response. The
cellular regulation machinery can adjust the binding rates and hence
the resulting spatio-temporal dynamics of signalling molecules
\cite{Alberts,CAM}. 

Therefore, in biological systems wide ranges of
$z$ and $\varphi$ occur and may have been selected by evolution. In the biophysical context the first passage time
problem studied here would
correspond to the association time of a signalling molecule with its
target. Our results show that changing $z$ and $\varphi$
can profoundly affect the association dynamics. In particular, our
results demonstrate that it is
possible to tune specific stages of the target search process, such as
delivery to the target from a distance or the hitting step from close
proximity. Conversely, our findings highlight the fact that the dynamics
cannot be quantified in terms of effective parameters alone,
e.g., with an average diffusion coefficient. Nor can the first arrival time
statistics be specified solely on the basis of MFPT concepts.

\section{Proofs and details of calculations}

In this section we describe details of the calculations and provide
proofs of the equations presented in the previous section. 

\subsection{Solution of the coupled mixed boundary value problem}

To solve Eq.~(\ref{FPE}) we first introduce auxiliary dimensionless coordinates
$\mathbf{x}'=\sqrt{k_1/D_1}\mathbf{x}$ and $t'=k_1t$ and
Laplace transform in time
$\tilde{\mathbf{p}}^{\mathrm{T}}=\hat{\mathcal{L}}[\mathbf{p}^{\mathrm{T}};t'\to
  s]$. Defining for convenience $z=k_2/k_1$ and $\varphi=D_2/D_1$,
we find that the components of $\tilde{\mathbf{p}}^{\mathrm{T}}$
obey
\begin{eqnarray}
\label{first}
(\nabla^2-1-s)\tilde{p}_1+z\tilde{p}_2=-\frac{w}{4\pi r_0^{'2}\sqrt{k_1D_1}}\delta(r'-r'_0)\\
(\varphi\nabla^2-z-s)\tilde{p}_2+\tilde{p}_1=-\frac{(1-w)}{4\pi r_0^{'2}\sqrt{k_1D_1}}\delta(r'-r'_0),
\label{second}
\end{eqnarray} 
where we take either $w=1$ or $w=0$, as any other solution is
obtained by linear superposition of these
solutions. Since we assume that initially the particle's position is uniformly
distributed over the surface of a sphere with radius $r_0$ (see
section 2), the boundary value problem in Eqs.~(\ref{first}) and
(\ref{second}) becomes
effectively 1-dimensional in the radial coordinate. Eqs.~(\ref{first}) and (\ref{second}) show that $\tilde{p}_{1,2}$ correspond
to the Green's functions of our coupled mixed boundary value problem. 
The general solution to the homogeneous coupled equations is obtained
by inserting Eq.~(\ref{second}) into Eq.~(\ref{first}) to obtain the 4th
order PDE for $\tilde{p}_1$
\begin{equation}
\label{fourth}
\left\{\varphi\nabla^4-[(\varphi+1)s+\varphi+z]\nabla^2+s(s+z+1)\right\}\tilde{p}_1=0.
\end{equation}
To solve it we make the standard ansatz
\begin{equation}
\nabla^2\tilde{p}_1=q\tilde{p}_1 
\label{ansatz}
\end{equation}
such that $q$ is the root of the
quadratic equation 
\begin{equation}
\varphi q^2-[(\varphi+1)s+\varphi+z]q+s(s+z+1)=0
\label{quadratic}
\end{equation}
or explicitly,
\begin{equation}
q_\pm(s)=\frac{1}{2\varphi}\left[(\varphi+1)s+\varphi+z\pm\sqrt{[(1-\varphi)s+(z-\varphi)]^2-4\varphi
    z}\right].
\label{root}
\end{equation}
The general solution of Eqs.~(\ref{first}) and (\ref{second}) for a 3-dimensional system with
spherical symmetry can now be written as 
\begin{eqnarray}
\label{partic1}
\tilde{p}_1(r',s)&=&r'^{-1}\left(C_1\mathrm{e}^{-\sqrt{q_+}r'}+C_2\mathrm{e}^{\sqrt{q_+}r'}+C_3\mathrm{e}^{-\sqrt{q_-}r'}+C_4\mathrm{e}^{\sqrt{q_-}r'}\right)\\
\tilde{p}_2(r',s)&=&\left(\frac{s+1-q_+}{r'z}\left[C_1\mathrm{e}^{-\sqrt{q_+}r'}+C_2\mathrm{e}^{\sqrt{q_+}r'}\right]\right.\nonumber\\
&&+\left.\frac{s+1-q_-}{r'z}\left[C_3\mathrm{e}^{-\sqrt{q_-}r'}+C_4\mathrm{e}^{\sqrt{q_-}r'}\right]\right),
\label{partic2}
\end{eqnarray}
where Eq.~(\ref{partic1}) is obtained as a solution of
Eq.~(\ref{ansatz}) and Eq.~(\ref{partic2}) is obtained by first inserting the solution (\ref{partic1})
into the homogeneous form of Eq.~(\ref{first}) and then solving for $\tilde{p}_2$. Moreover, $C_1$ to $C_4$ are constants determined by the boundary conditions
in Eq.~(\ref{BC}), and the continuity and jump discontinuity of the
Green's functions in Eqs.~(\ref{first}) and (\ref{second}). These lead to two
inhomogeneous systems of 8
linear equations with 8 unknowns, $C_1$ to $C_4$ for $r\le r_0$ and
$C_5$ to $C_8$ for $r> r_0$, for each of the cases $w=0$ and $w=1$,
respectively. These are in turn solved by Cramer's rule. We omit these
calculations as they are tedious but straight forward. 

The Laplace transformed FPT density $\tilde{\wp}(s)$ in the
dimensionless units introduced in section 2.1 is obtained from the flux into the
absorbing boundary (i.e. from the Laplace transform of
Eq.~(\ref{flux})) and in the final form reads for the recognition and
non-recognition initial condition, respectively,
\begin{eqnarray}
\label{LFLux1}
\tilde{\wp}_1(s)=\left(\frac{x_a}{x_0}\right)\frac{Q_2(s)\Delta_1(s,x_0)\mathcal{D}_2(s,x_a)-Q_1(s)\Delta_2(s,x_0)\mathcal{D}_1(s,x_a)}{Q_2(s)\Delta_1(s,x_a)\mathcal{D}_2(s,x_a)-Q_1(s)\Delta_2(s,x_a)\mathcal{D}_1(s,x_a)},\\
\tilde{\wp}_2(s)=\left(\frac{x_a}{x_0}\right)\frac{\varphi^{-1}[\Delta_2(s,x_0)\mathcal{D}_1(s,x_a)-\Delta_1(s,x_0)\mathcal{D}_2(s,x_a)]}{Q_2(s)\Delta_1(s,x_a)\mathcal{D}_2(s,x_a)-Q_1(s)\Delta_2(s,x_a)\mathcal{D}_1(s,x_a)}.
\label{LFLux2}
\end{eqnarray}
Here we introduced the auxiliary functions
$Q_{1,2}(s)=(s+k_1-k_1q_{+,-}(s))/(k_1z)$ as well as
\begin{eqnarray}
\label{aux_L1}
\mathcal{D}_{1,2}(s,y)&=&(1-y)\cosh[\sqrt{k_1q_{+,-}(s)}(1-y)]\nonumber\\
&+&\frac{[yk_1q_{+,-}(s)-1]\sinh[\sqrt{k_1q_{+,-}(s)}(1-y)]}{\sqrt{k_1q_{+,-}(s)}}\\
\Delta_{1,2}(s,y)&=&\cosh[\sqrt{k_1q_{+,-}(s)}(1-y)]-\frac{\sinh[\sqrt{k_1q_{+,-}(s)}(1-y)]}{\sqrt{k_1q_{+,-}(s)}},
\label{aux_L2}
\end{eqnarray}  
where we always take the first or second index on both sides,
respectively. Note that here we already back-transformed the auxiliary
coordinates $\mathbf{x}'\to\mathbf{x}$ and $s\to s/k_1$. Obviously, $\overline{\mathcal{D}}_1(y)=\mathcal{D}_{1}(0,y)$
and $\overline{\Delta}_1(y)=\Delta_1(0,y)$ (see
Eqs.~(\ref{aux1}) and (\ref{aux2})). Note that $\tilde{\wp}_{1,2}(s)$ has a removable
singularity at $s=0$, therefore we re-define the analytic function
$\tilde{\wp}_{1,2}(s)$ at $s=0$ as
$\tilde{\wp}_{1,2}(0)\equiv\lim_{s\to 0}\tilde{\wp}_{1,2}(s)$.

\subsection{Mean first passage times}

Proving Eqs.~(\ref{MFPT1}) and (\ref{MFPT2}) is henceforth
easy, and is carried out by taking the derivative of Eqs.~(\ref{LFLux1}) and (\ref{LFLux2}) 
\begin{equation}
\label{MFPTp}
\langle t_{x_a}(x_0)\rangle_{1,2} =
\left.-\frac{\partial \tilde{\wp}_{1,2}(s)}{\partial s}\right|_{s=0}.
\end{equation}
Noticing that $\overline{\mathcal{D}}_2(y)=0$,
$\overline{\Delta}_2(y)=y$ as well as $q_-(0)=0$,
$q_+(0)=k_1(1+z/\varphi)$, and finally
$q_+'(0)=(z/\varphi+\varphi)/(z+\varphi)$, $q_-'(0)=(1+z)/(z+\varphi)$, 
and performing some elementary algebraic manipulations  already
completes the proof of Eqs.~(\ref{MFPT1}) and (\ref{MFPT2})$_\square$

\subsection{Inverse Laplace transform of $\tilde{\wp}(s)$}

\emph{a) Justification of assumptions (i) to (iii) made in section
3.2}\\
\\
\noindent Note that the analytic function $\tilde{\wp}(s)$ defined in section
4.1 is regular at $s=0$, has no branch points on the negative real
axis (hence justifying assumption (ii) in Section
3.2)  and allows a moment expansion
$\tilde{\wp}(s)=\sum_{n=0}^{\infty}(-s)^n\langle t ^n\rangle/n!$ converging for
$\mathrm{Re}(s)<\lambda_0$, where $-\lambda_0\in \mathbb{R}$ is the pole of
$\tilde{\wp}(s)$ closest to the origin \cite{Godec_PRX}. This also implies that all moments of
$\wp(t)$ are finite, which justifies assumption (i) in Section
3.2 \cite{Godec_PRX}. Moreover, the moments $\langle t^n\rangle$ are obtained
recursively from Taylor
coefficients of the series of the numerator and denominator of
Eqs.~(\ref{LFLux1}) and (\ref{LFLux2}),
$\sum_{k=0}^{\infty}u_{1,2}^{(k)}(0)s^k/k!$ and
$\sum_{k=0}^{\infty}v^{(k)}(0)s^k/k!$, respectively,
\begin{equation}
\langle t^n\rangle=(-1)^n\frac{u^{(n)}(0)}{v^{(0)}(0)}-\sum_{k=1}^{n}(-1)^k{{n}\choose{k}}\frac{v^{(k)}(0)}{v^{(0)}(0)}\langle t^{n-k}\rangle.
\end{equation}
 Explicitly, the
coefficients $u_1^{(k)}(0)$ of the numerator read
\begin{eqnarray}
u_1^{(n)}(0)=n!\left(\frac{1+\varphi}{k_1(z+\varphi)}\right)^n\sum_{k=0}^{\infty}\left(\frac{k_1(z+\varphi)}{2\varphi}\right)^k\nonumber\\
\times\sum_{l=0}^{k}\frac{(1-x_0)^{2(k-l)}(1-x_a)^{2l+1}[2(k-l)+x_a]}{(2[k-l]+1)!(2l+1)!}\nonumber\\
\times\left[\mathcal{S}_1(k,l,n)+\mathcal{S}_2(k,l,n)+\mathcal{S}_3(k,l,n)+\mathcal{S}_4(k,l,n)-\mathcal{S}_5(k,l,n)\right]
\label{Tnumerator}
\end{eqnarray}
where the functions $\mathcal{S}_1$ to $\mathcal{S}_5$ are defined as 
\begin{eqnarray}
\label{S1}
\mathcal{S}_1(k,l,n)&=&\frac{2l(1-x_a)-x_a}{x_0z}
\sum_{i=0}^{\lfloor\frac{k-l}{2}\rfloor\wedge\lfloor\frac{l}{2}\rfloor}\Theta(k-n)\nonumber\\
&&\times\Theta(k-2i-n)\Theta(2i-n)W_{k-l,l,i}\Xi_{n,k-l,l,i},\\
\label{S2}
\mathcal{S}_2(k,l,n)&=&\frac{2l(1-x_a)+x_a(k_1-1)}{x_0k_1z}
\sum_{i=0}^{\lfloor\frac{k-l}{2}\rfloor\wedge\lfloor\frac{l+1}{2}\rfloor}\Theta(k+1-n)\nonumber\\&&\times\Theta(k+1-2i-n)\Theta(2i-n)W_{k-l,l+1,i}\Xi_{n,k-l,l+1,i},\\
\label{S3}
\mathcal{S}_3(k,l,n)&=&\frac{2l(1-x_a)-x_a}{x_0k_1z}
\sum_{i=0}^{\lfloor\frac{k-l}{2}\rfloor\wedge\lfloor\frac{l}{2}\rfloor}\Theta(k+1-n)\nonumber\\&&\times\Theta(k+1-2i-n)\Theta(2i+1-n)W_{k-l,l,i}\Xi_{n-1,k-l,l,i},\\
\label{S4}
\mathcal{S}_4(k,l,n)&=&\frac{x_a}{x_0k_1z}\sum_{i=0}^{\lfloor\frac{k-l}{2}\rfloor\wedge\lfloor\frac{l+1}{2}\rfloor}\Theta(k+2-n)\nonumber\\
&&\times\Theta(k+2-2i-n)\Theta(2i+1-n)W_{k-l,l+1,i}\Xi_{n-1,k-l,l+1,i},\\
\label{S5}
\mathcal{S}_5(k,l,n)&=&\frac{x_a}{x_0k_1z}\sum_{i=0}^{\lfloor\frac{k-l}{2}\rfloor\wedge\lfloor\frac{l+2}{2}\rfloor}\Theta(k+2-n)\nonumber\\
&&\times\Theta(k+2-2i-n)\Theta(2i-n)W_{k-l,l+2,i}\Xi_{n,k-l,l+2,i},
\end{eqnarray}
where $\Theta(n)$ is the discrete Heaviside step function, $\lfloor x \rfloor$ is the floor function, $x \wedge
y\equiv\mathrm{min}(x,y)$ and
\begin{eqnarray}
\label{W}
W_{p,q,i}=\sum_{j=0}^i&&\left\{\Theta(\lfloor
  q/2-1\rfloor-j)\Theta(\lfloor p/2\rfloor-i){{p}\choose{2i}}{{q}\choose{2j+1}}\right.\nonumber\\
&&\left.+\Theta(\lfloor q/2\rfloor-j)\Theta(\lfloor p/2-1\rfloor-i){{p}\choose{2j+1}}{{q}\choose{2i}}\right\},
\end{eqnarray}
and where we also introduced
\begin{eqnarray}
\label{Xi}
\Xi_{k,p,q,i}&=&\sum_{m=0}^k\Theta(p+q-2i+m-k)\Theta(2i-m){{p+q-2i}\choose{k-m}}{{2i}\choose{m}}\nonumber\\
&&\times\left[\frac{(1-\varphi)(z+\varphi)}{(1+\varphi)(z-\varphi)}\right]^m\left[\frac{(z-\varphi)^2}{k_1z\varphi}\right]^{\lfloor
  (k-m)/2\rfloor}{{2i}\choose{\lfloor
    (k-m)/2\rfloor}}\nonumber\\
&&\times{_2F_1}\left\{1,\lfloor
  (k-m)/2\rfloor-2i,1+\lfloor
  (k-m)/2\rfloor;-\frac{(z-\varphi)^2}{k_1z\varphi}\right\}.
\end{eqnarray}
Finally, $_2F_1\{i,j,k;z\}$ denotes the Gauss hypergeometric
function. The coefficients $v^{(k)}(0)$ of the denominator are
obtained by replacing $x_0$ with $x_a$. 

For the scenario of starting in the non-recognition mode the Taylor
series of the numerator is simpler and the coefficients read
\begin{eqnarray}
u_2^{(n)}(0)=-\frac{n!}{\varphi}\left(\frac{1+\varphi}{k_1(z+\varphi)}\right)^n\sum_{k=0}^{\infty}\left(\frac{k_1(z+\varphi)}{2\varphi}\right)^k\nonumber\\
\times\sum_{l=0}^{k}\frac{(1-x_0)^{2(k-l)}(1-x_a)^{2l+1}[2(k-l)+x_a]}{(2[k-l]+1)!(2l+1)!}\nonumber\\
\times\Big[k_1z\mathcal{S}_1(k,l,n)+\frac{x_a}{x_0}\Theta(k+1-n)\sum_{i=0}^{\lfloor\frac{k-l}{2}\rfloor\wedge\lfloor\frac{l+1}{2}\rfloor}\Theta(k+1-2i-n)\nonumber\\
\times\Theta(2i-n)W_{k-l,l+1,i}\Xi_{n,k-l,l+1,i}\Big].
\label{Tnumerator2}
\end{eqnarray}
The calculation leading to the Taylor series (\ref{Tnumerator}) and (\ref{Tnumerator2}) is essentially straightforward and amounts to combining the respective Taylor series of the
individual functions occurring in Eqs.~(\ref{LFLux1}) and
(\ref{LFLux2}) and carefully performing a sequence of changes of the
order of summations thereby bringing the summation over powers of $s$
to the outermost sum. The numerous step functions $\Theta(x)$ in the
expressions for the coefficients are merely a consequence of the
preservation of the domain of summation upon changing the order in which
they are carried out. 

The coefficients $u_1^{(k)}(0),u_2^{(k)}(0)$ and $v^{(k)}(0)$ are hence
given in the form of convergent series and it is not difficult to check
(e.g., using \textsf{Mathematica}) that
$\lim_{n\to\infty}u_1^{(n)}(0)/v^{(n)}(0)=0$ and
$\lim_{n\to\infty}u_2^{(n)}(0)/v^{(n)}(0)=0$, thereby justifying the
assumption (iii) of Section 3.2., that is
$\lim_{n\to\infty}u^{(n)}(0)/v^{(n)}(0)<\infty$. Summing up this now
justifies
assumptions (i) to (iii) in Section 3.2, i.e., the necessary conditions for the
validity of Eqs.~(\ref{full}) to (\ref{matrix}) \cite{Godec_PRX}.\\  
\\
\noindent \emph{b) Asymptotic inversion of the Laplace transform}\\
\\
\noindent Since $\tilde{\wp}(s)$ has no branch points we can invert it
using Cauchy's theorem
\begin{equation}
\wp(t)\sim \lim_{s\to-\lambda_0}[(s+\lambda_0)\tilde{\wp}(s)\mathrm{e}^{st}],
\label{Cauchy}
\end{equation}
where the contour used to evaluate the residue is chosen as to
enclose $-\lambda_0$ such that $\mathbb{R}(s)<\lambda_0$. A rigorous
solution to this problem, i.e. determining $\lambda_0$ and evaluating
the residue in Eq.~(\ref{Cauchy}), was obtained recently under the
assumptions (i) to (iii) in Section 3.2 \cite{Godec_PRX}. A detailed
proof is given in Ref. \cite{Godec_PRX}. Here we
merely state the result, which has the form of
Eqs.(\ref{full}) to (\ref{matrix}).

To proceed towards our central result Eq.~(\ref{FPTas}) we note that
the first terms of the series (\ref{full}) are 
\begin{equation}
\lambda_0(x_a)=\frac{v^{(0)}(x_a)}{v^{(1)}(x_a)}\left(1+\frac{v^{(0)}(x_a)}{2}\frac{v^{(2)}(x_a)}{v^{(1)}(x_a)^2}[1+\mathcal{O}]\right),
\end{equation}
where $\frac{v^{(0)}(x_a)}{2}\frac{v^{(2)}(x_a)}{v^{(1)}(x_a)^2}$ is
of the order of $x_a$ and moreover, $\mathcal{O}$ is also of the order
of $x_a$. This can be seen either by computing the respective
derivatives explicitly or from the Taylor coefficients in
Eq.~(\ref{Tnumerator}) by making the replacement $x_0\to
x_a$. Therefore all correction terms vanish in the limit $x_a\to 0$,
and  Eqs.~(\ref{full}) and  (\ref{pref}) both fully converge already with
the first term, which completes the proof of Eq.~(\ref{FPTas})$_\square$ 

\section{Conclusion}

Our results highlight the complex character of the first passage time
statistics of Markov additive processes. While it appears to be a
common feature of non-recurrent Markov processes that the first passage
time asymptotics can be fully reconstructed from the corresponding mean first passage
time \cite{GodSciRep,Olivier}, we here showed that this is not the case for Markov
additive processes. The present results on  a Markovian sum of two
perfectly non-recurrent Bessel processes establish rigorously the non-trivial
connection between mean first passage
times and long time first passage asymptotics. In addition, we also
obtained results for the case of transient immobilisation, i.e., the transitioning into
an immobile phase.

The results of this paper are
important in a broader context, as most existing studies of the
first passage behaviour of Markov additive processes are limited to the
analysis of mean first passage times
\cite{Holcman,OlivierAct,gene}. Moreover, our results
also demonstrate that the first passage behaviour of Markov additive
processes in general cannot be adequately captured by effective quantities such
as the effective diffusion coefficient. This is important if one would
attempt to develop effective medium or averaging type approximations.

The exact Laplace inversion formula presented
in this paper (Eqs.~(\ref{full}) to (\ref{matrix})) will be useful in
various problems, as it reduces the problem of deriving first passage
asymptotics to the much simpler problem of finding the Laplace transform of the
first passage time density. It will also be very useful for developing
singular perturbation results, such as the small target limit studied here.

The present results can be extended in numerous ways. For instance, a
straightforward extension would be to include more internal states,
or to combine diffusive and advective states such as in the
intermittent search model \cite{OlivierAct} or in the presence of so-called
cytoplasmic streaming in cells \cite{streaming}.
One could also take into account the spatial heterogeneity of diffusion
coefficients \cite{GodPRE,GodSciRep,heterogen}, spatial or energetic disorder
\cite{disorder} or consider a more complex fluctuating environment \cite{GodNJP}. Similarly, one could address the role of anomalous
diffusion, such as observed in the motion of proteins and submicron
objects in the cell cytoplasm \cite{anomalous}.

\ack

We thank A. Cherstvy for stimulating discussions. AG acknowledges funding through an Alexander von Humboldt Fellowship
and ARRS project Z1-7296.


\end{document}